\documentclass[10pt,final,journal,letterpaper,twoside]{IEEEtran}
\renewcommand{\baselinestretch}{0.98}
\usepackage[usenames,dvipsnames,svgnames,table]{xcolor}
\usepackage{booktabs,bbding,nicefrac}
\usepackage{listings}
\usepackage{units}
\usepackage{amssymb,dsfont,multirow,cite}
\usepackage[framed,numbered,autolinebreaks,useliterate]{mcode}
\ifCLASSINFOpdf
   \usepackage[pdftex]{graphicx}
   \graphicspath{{figures/}}
   \DeclareGraphicsExtensions{.pdf,.jpeg,.png}
\else
  \usepackage[dvips]{graphicx}
   \graphicspath{{figures/eps/}}
   \DeclareGraphicsExtensions{.eps}
\fi
\usepackage{amsmath}
\ifCLASSOPTIONcompsoc
  \usepackage[caption=false,font=normalsize,labelfont=sf,textfont=sf]{subfig}
\else
  \usepackage[caption=false,font=footnotesize]{subfig}
\fi
\usepackage{fixltx2e,dblfloatfix,color,soul}
\usepackage[colorlinks,allcolors=blue]{hyperref}
\begin{document}
\title{Transmission Ratio Design for Electric Vehicles\\ via Analytical Modeling and Optimization}
\author{Theo Hofman and Mauro Salazar
\thanks{Dr. Theo Hofman and Dr. Mauro Salazar are with the Control Systems Technology group, Dep. of Mechanical Engineering, Eindhoven University of Technology, The Netherlands, (e-mail: \texttt{\{t.hofman,.m.r.u.salazar\}@tue.nl}).}}%
\maketitle
\begin{abstract}
In this paper we present an effective analytical modeling approach for the design of the transmission of electric vehicles.
Specifically, we first devise an analytical loss model for an electric machine and show that it can be accurately fitted by only sampling three points from the original motor map.
Second, we leverage this model to derive the optimal transmission ratio as a function of the wheels' speed and torque, and use it to optimize the transmission ratio.
Finally, we showcase our analytical approach with a real-world case-study comparing two different transmission technologies on a BMW~i3: a fixed-gear transmission (FGT) and a continuously variable transmission (CVT). Our results show that even for e-machines intentionally designed for a FGT, the implementation of a CVT can significantly improve their operational efficiency by more than 3\%.
The provided model will ultimately bridge the gap in understanding how to efficiently specify the e-machine and the transmission technology in an integrated fashion, and enable to effectively compare single- and multi-speed-based electric powertrains.
\end{abstract}
\begin{IEEEkeywords}
Electric vehicles; powertrains; electric machines; transmissions; automated transmissions; continuously variable transmission; optimization
\end{IEEEkeywords}
\IEEEpeerreviewmaketitle

\section{Introduction}\label{sec:intro}
\IEEEPARstart{E}{lectric} vehicles are being rapidly introduced into the market causing a large market shift in the mid- and long-term from the classical combustion-engine-driven towards all-electric-driven powertrains for a range of applications, e.g., passenger cars, electric buses and trucks \cite{Ver18, Ver19}, and even race cars~\cite{FIA,BorsboomFahdzyanaEtAl2020}. These new powertrain types have intrinsically a 2-3 times higher efficiency compared to combustion-engine-based powertrains. Yet, they do show potential for further improvement that could be gathered by investigating different electrical powertrain topologies or layouts, electric machine designs in combination with the power electronics and battery pack optimization~\cite{Verbruggen_2020}.
The electric machine type alone can have a significant impact on the vehicle energy usage~\cite{Zhu17}, which may vary between 11-14\% for (8) equally sized machines (Toyota Prius motor as reference) ranging from asynchronous induction machines (AIM) to permanent magnet synchronous machines (PMSM), respectively.
Some works focus on the design of the e-machine depending on the operation conditions~\cite{Choi13}. The design of an e-machine is a highly complex process due to its multi-objective and nonlinear nature of the mathematical design problem~\cite{Lei17}, which can result in long computation times~\cite{Zuu2019}.
To avoid a large asymmetric design optimization process of the powertrain, some researchers use scaling laws or descriptive analytic models to capture the energy efficiency characteristics to speed up the scaling process~\cite{Sti15, Zhao16}.
\begin{figure}[!t]
\centering
\subfloat[FGT: 1-speed]{\includegraphics[width= 0.4\columnwidth]{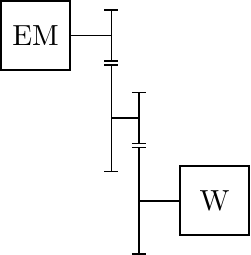}}
\subfloat[AMT: 2-speed]{\includegraphics[width= 0.4\columnwidth]{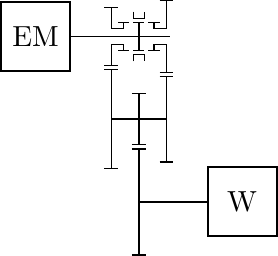}}\\
\subfloat[DCT: 2-speed]{\includegraphics[width= 0.4\columnwidth]{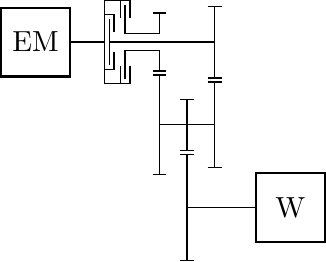}}
\subfloat[CVT: variable]{\includegraphics[width= 0.4\columnwidth]{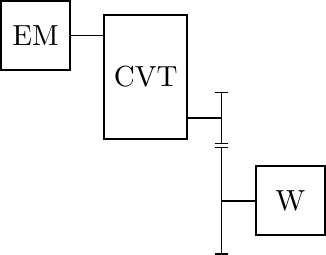}}
 \caption{Electric powertrain topologies (from \cite{Hup19}) used in different studies \cite{Gun13, Rua16,Gao15,Guo17,Zhou19,Rua18,Tia18,Pin14}: fixed gear transmission (FGT) (a), automated manual transmission (AMT) (b), dual clutch transmission (DCT) (c), continuously variable transmission (CVT) (d)}\label{fig:fig1}
\end{figure}

\subsection{Transmission Design: from Single-speed to Multi-speed}
The gear ratio design for a fixed-gear transmission (FGT) with an electric machine is mainly found as a result of a tradeoff between the desired drivability requirements, e.g., acceleration or uphill driving and top speed specifications. For acceleration, in particular, high torque (and power) is required, while at the same time sufficient torque at a relative high vehicle speed without overspeeding the machine is needed.
Using the machine at high vehicle speeds requires a lower gear ratio value, which inherently compromises the desired torque amplification at low speeds and causes a significant current (and torque) overload during vehicle launch.
Without an improved or advanced cooling system while repetitively performing full accelerations, the maximum amount of full accelerations without compromising drivability is limited. Thereby, even the power output may be limited afterwards, to avoid overheating the powertrain.

A multi-speed transmission reduces the need for overloading the electric machine, while at the same time uncompromising the maximum achievable vehicle top speed. It can not only reduce the nominal torque, but also the usage of active materials and maximum machine speed specifications, which partly compensates the additional shift actuation losses with lower mechanical losses due to bearings and internal friction. Since electric machines have a higher speed range compared to engines, using a multi-speed for an electric vehicle reduces the required ratio coverage of the transmission around 3-4 times, resulting in a compact design~\cite{Hup19,Gun13}.

\subsection{Contribution and Outline of the Paper}
Studies related to new continuously variable transmission (CVT) design for EVs show that the torque and speed specification can be reduced by -61\% (from \unit[213]{Nm} to \unit[83]{Nm}) and -25\% (from \unit[16]{krpm} to \unit[12]{krpm}), respectively. The physical machine size and battery pack can be reduced by -45\% and -12\%, respectively, and improves the overall efficiency in the order of 6\% lower than \unitfrac[10]{kWh}{100km} on the WLTC for a vehicle of \unit[1415]{kg} \cite{Hup19}.
In recent years, more research work can be found on the control and design of multi-speeds for electric powertrains in passenger cars \cite{Gun13, Rua16,Gao15,Guo17,Zhou19,Rua18,Tia18,Pin14,Verbruggen_2020,VerbruggenSalazarEtAl2019} and Fig.~\ref{fig:fig1} shows some of the investigated transmission technologies. However, the design of the machine is mostly decoupled from the transmission design and done for only one particular machine technology. This strongly limits the optimality in design, hence our attempt, in this study, to understand the coupling between the machine's efficiency and optimal speed ratio design for two different transmission technologies: a CVT and a FGT. We leverage complex analysis by introducing an analytical model that captures with sufficient accuracy the machine's efficiency. This model requires \textit{only} a maximum of three fitted machine operation points to fully describe the machine's efficiency as a function of torque and speed. Consequently, an explicit solution for the optimal operation line can be derived and used to find the optimal speed ratio values for arbitrary traction-torque and speed demands at the wheels.
In this analysis, a highly efficient PMSM machine from BMW will be used, with a peak efficiency around 97\%~\cite{Mer14}.

The outcome of the study may result in new design specifications for electric machine engineers and enable insights into the strong coupling of machine and transmission design. To limit our research we focus on passenger car applications and restrict the study to the topology shown in Fig.~\ref{fig:fig1}. It should be noted that extending the study to other applications and the vast design space of alternative powertrain topologies would require additional work that we leave to future research endeavors.

The remainder of this paper is structured as follows: First, the analytical loss model will be introduced in Section~\ref{sec:modeling}. Thereby, the found model will be evaluated using a PMSM machine map in Section~\ref{sec:recon}. Efficiencies will be calculated leveraging the proposed analytical model, original data map model for a CVT and a fixed-gear transmission. The optimal gear ratio design is derived and evaluated in Section~\ref{sec:gear}. Finally, the conclusions will be summarized in Section~\ref{sec:conc}.

\section{Electric Machine Modeling and Operation}\label{sec:modeling}
This section presents the analytical loss model that we will leverage to derive an analytical expression for the efficiency-optimal machine operation in terms of torque and speed, which we will validate with a data map model of a PMSM.
Thereafter, we will implement the results to calculate the optimal transmission ratio values. Finally, the energy usage under dynamic driving conditions (WLTC) using the original data-based and the analytical efficiency map for a CVT and a FGT with optimized speed ratios will be compared.

\subsection{Descriptive Analytical Loss Model}
The power losses $P_\mathrm{loss}$ of an electrical machine can be expressed as a function of torque $\tau_\mathrm{m}$ and speed $\omega_\mathrm{m}$ as
\begin{small}
\begin{equation} \label{eq:eqloss}
P_\mathrm{loss}(\tau_\mathrm{m}, \omega_\mathrm{m}) = \sum_{m,n} c_{m n} \, \tau^m_\mathrm{m} \, \omega^n_\mathrm{m},
\end{equation}
\end{small}
\!\!where the exponents $m$ and $n$ are non-negative integers, and the constants $c_{m n}$ are non-negative~\cite{Mah17}. It has been found that using coefficients up to the order of two or three is sufficiently accurate to describe the different electrical and mechanical loss terms related to, e.g., copper ($\propto \tau_\mathrm{m}^2$), iron ($\propto \omega_\mathrm{m}^2$), magnetic, windage ($\propto \omega_\mathrm{m}^3$) or friction losses. We rewrite~\eqref{eq:eqloss} as follows for $m$ and $n$ up to $2$ so that it becomes a convex function w.r.t.\ torque:
\begin{small}
\begin{equation} \label{eq:eqloss2}
P_\mathrm{loss}(\tau_\mathrm{m}, \omega_\mathrm{m}) = \begin{bmatrix}
 \alpha_1(\omega_\mathrm{m}) & \alpha_2(\omega_\mathrm{m}) & \alpha_3(\omega_\mathrm{m})
 \end{bmatrix}
 \,
  \begin{bmatrix}
 1 \\
\tau_\mathrm{m} \\
\tau_\mathrm{m}^2
 \end{bmatrix}
\end{equation}
\end{small}
\!\!with
\begin{small}
\begin{equation} \label{eq:eqalpha}
\begin{bmatrix}
 \alpha_3(\omega_\mathrm{m}) \\
 \alpha_2(\omega_\mathrm{m}) \\
 \alpha_1(\omega_\mathrm{m})
 \end{bmatrix}
 =
 \begin{bmatrix}
 c_{2 0} & c_{2 1} & c_{2 2} \\
 c_{1 0} & c_{1 1} & c_{1 2} \\
 c_{0 0} & c_{0 1} & c_{0 2}
 \end{bmatrix}
 \,
 \begin{bmatrix}
 1 \\
\omega_\mathrm{m} \\
\omega_\mathrm{m}^2
 \end{bmatrix}.
 \end{equation}
 \end{small}
\!\!In the literature, different models coexist, e.g., by assuming only $\{c_{0 0},c_{0 1},c_{0 2},c_{1 1},c_{2 0}\} \in \mathds{R}^+$ and the other coefficients to be zero
as in~\cite{Zha17} for a modified two-degree model (evaluated using an AIM and a PMSM machine), or even only $\{c_{2 0}, c_{0 1}, c_{0 3}, c_{0 0}\} \in \mathds{R}^+$ for a three-degree model in \cite{HOF2017} for a PMSM. In~\cite{Zha17} and in~\cite{Pur13}, the losses are scaled by scaling the fit coefficients $c_{m n}$ with the rated (or peak) torque, denoted as $\overline{\tau}_\mathrm{m}$, and speed, denoted as $\omega_\mathrm{m,0}$, or with the rated torque alone, respectively.

\subsection{Optimal Machine Operation: Analytical Solution}
The machine efficiency is described as the ratio of the mechanical output power $P_\mathrm{m}=\tau_\mathrm{m} \, \omega_\mathrm{m}$ with the AC machine input power from the inverter $P_\mathrm{ac}$ defined as the sum of the output power and the machine losses, i.e.,
\begin{small}
\begin{equation}\label{eq:eff}
 \eta_\mathrm{m}(\tau_\mathrm{m},\omega_\mathrm{m}) = \frac{P_\mathrm{m}}{P_\mathrm{ac}} = \frac{\tau_\mathrm{m} \, \omega_\mathrm{m}}{\tau_\mathrm{m} \, \omega_\mathrm{m} + P_\mathrm{loss}}.
\end{equation}
\end{small}
\!\!The  maximum efficiency machine torque can be found from
\begin{small}
\begin{equation}\label{eq:par1}
\frac{\partial \eta_\mathrm{m}}{\partial \tau_\mathrm{m}} = 0,
\end{equation}
\end{small}
\!\!which results in the optimal machine torque after substitution of \eqref{eq:eqloss2} in \eqref{eq:eff} and solving for \eqref{eq:par1} as
\begin{small}
\begin{equation}\label{eq:opttrq}
\tau^*_\mathrm{m}(\omega_\mathrm{m}) = \sqrt{\frac{\alpha_1(\omega_\mathrm{m})}{\alpha_3(\omega_\mathrm{m})}} =  \sqrt{\beta(\omega_\mathrm{m})}.
\end{equation}
\end{small}
\!\!In particular, for a modified two-degree loss model with $\{c_{0 0},c_{0 1},c_{0 2},c_{1 1},c_{2 0}\} \in \mathds{R}^+$ as in \cite{Zha17} the speed-dependent term $\beta(.)$ becomes
\begin{small}
\begin{equation}\label{eq:beta}
\beta(\omega_\mathrm{m}) = d_{0 0} + d_{0 1} \, \omega_\mathrm{m} + d_{0 2} \, \omega_\mathrm{m}^2
\end{equation}
\end{small}
\!\!with coefficients
\begin{small}
\begin{equation}\label{eq:d}
d_{0 0} = c_{0 0}/c_{2 0},~d_{0 1} = c_{0 1}/c_{2 0},~d_{0 2} = c_{0 2}/c_{2 0}.
\end{equation}
\end{small}
\!\!Therefore, the optimal machine efficiency as a function of speed can be expressed as
\begin{small}
\begin{equation}\label{eq:effred}
\eta_\mathrm{m}^*(\omega_\mathrm{m}) = \frac{\omega_\mathrm{m}}{\omega_\mathrm{m} + 2 \, \alpha_3 \, \tau^*_\mathrm{m}(\omega_\mathrm{m}) + \alpha_2(\omega_\mathrm{m})},
\end{equation}
\end{small}
\!\!and solving the partial derivative of the efficiency to speed as
\begin{small}
\begin{equation}\label{eq:par}
\frac{\partial \eta_\mathrm{m}^*}{\partial \omega_\mathrm{m}} = 0
\end{equation}
\end{small}
\!\!results in the optimal machine speed $\omega_\mathrm{m}^*$ given by~\eqref{eq:optspd}.
\begin{figure*}[!ht]
\footnotesize
\begin{tiny}
\begin{equation}\label{eq:optspd}
\omega^*_\mathrm{m} = \frac{2\,c_{2 0}\,\eta_\mathrm{m}^*\,\left(\sqrt{d_{0 0}\,{c_{1 1}}^2\,{\eta_\mathrm{m}^*}^2+2\,d_{0 0}\,c_{1 1}\,{\eta_\mathrm{m}^*}^2-2\,d_{0 0}\,c_{1 1}\,\eta_\mathrm{m}^*+{c_{2 0}}^2\,{d_{0 1}}^2\,{\eta_\mathrm{m}^*}^2-4\,d_{0 0}\,d_{0 2}\,{c_{2 0}}^2\,{\eta_\mathrm{m}^*}^2+d_{0 0}\,{\eta_\mathrm{m}^*}^2-2\,d_{0 0}\,\eta_\mathrm{m}^*+d_{0 0}}+c_{2 0}\,d_{0 1}\,\eta_\mathrm{m}^*\right)}{{c_{1 1}}^2\,{\eta_\mathrm{m}^*}^2+2\,c_{1 1}\,{\eta_\mathrm{m}^*}^2-2\,c_{1 1}\,\eta_\mathrm{m}^*-4\,d_{0 2}\,{c_{2 0}}^2\,{\eta_\mathrm{m}^*}^2+{\eta_\mathrm{m}^*}^2-2\,\eta_\mathrm{m}^*+1} \end{equation}
\begin{equation} \label{eq:optgear}
\gamma^*(\tau_\mathrm{t}, \omega_\mathrm{t}) = \frac{\sqrt{12 \, \sigma_4 \sqrt{\sigma_1} - \sigma_5^2 \, \sqrt{\sigma_1} - 9 \, \sigma_2^{2/3} \, \sqrt{\sigma_1} + 3 \, \sqrt{6} \, \sigma_6 \sqrt{27 \, \sigma_6^2 + 72 \, \sigma_5 \, \sigma_4 + 3 \, \sqrt{3} \, \sigma_3 + 2 \, \sigma_5^3} - 12 \, \sigma_5 \, \sigma_2^{1/3} \, \sqrt{\sigma_1}}}{6 \, \sigma_2^{1/6} \, \sigma_1^{1/4}} - \frac{d_{0 1}}{4 \, d_{0 2} \, \omega_\mathrm{t}} - \frac{\sqrt{\sigma_1}}{6 \, \sigma_2^{1/6}}
\end{equation}
\end{tiny}
\hrulefill
\vspace*{4pt}
\end{figure*}
\subsection{Machine Efficiency Map Reconstruction}\label{sec:recon}
To reconstruct an arbitrary efficiency map we proceed as follows: First, three optimal torque values at corresponding speeds, denoted as \emph{design points}, need to be defined (e.g., starting at the origin). Second, the three fit coefficients for $d_{i j}$ needed in~\eqref{eq:beta} are solved via~\eqref{eq:beta} evaluated at the three design points $(\omega_{\mathrm{m},i}^*,\tau_{\mathrm{m},i}^*)$.
It should be noted that there is a limit on the location options for the second torque-speed combination, since \eqref{eq:opttrq} should always result in real roots, i.e., \mbox{$d_{0 1}^2 - 4~d_{0 0}~d_{0 2} > 0$}. Finally, given the information on the efficiency at these optimal points obtained via~\eqref{eq:effred}, only two efficiencies are needed where the machine output power is larger than zero.
This way, we can obtain the values for the coefficients $c_{2 0}$ and $c_{1 1}$ via~\eqref{eq:effred}.
Using the values for $c_{2 0}$ and $c_{1 1}$, the values for $c_{0 0}$, $c_{0 1}$, and $c_{0 2}$ follow naturally from~\eqref{eq:d}.

Using an experimental example based on a highly-efficient PMSM, we now highlight the effectiveness of the fitting process we just outlined for the proposed analytical model:
\begin{enumerate}
\item We assume that the optimal operation line starts at the origin and set our first design point there: \mbox{$(\tau_{\mathrm{m},1}^*,\omega_{\mathrm{m},1}^*) = (\unit[0]{Nm},\unit[0]{rpm})$};
\item We find the torque and speed combination where the machine efficiency is maximized for the third design point: \mbox{$(\tau_{\mathrm{m},3}^*,\omega_{\mathrm{m},3}^*) = \arg\max_{\tau,\omega} \eta_\mathrm{m}$};
\item We iterate on the second design point $(\tau_{\mathrm{m},2}^*,\omega_{\mathrm{m},2}^*)$ in order to minimize the efficiency error of the analytical model at the design points.
\end{enumerate}
Critically, two design points, namely, the first and the third one, are straightforward to find. Therefore, the designer only needs to focus on the location of the second design point as illustrated below.
\begin{figure}[t]
\centering
\subfloat[Original map $\mathcal{M}_0$ versus the extracted data map $\mathcal{M}_1$.]{\includegraphics[clip = true, trim = 0cm 0cm 0cm 0cm,width= 1\columnwidth]{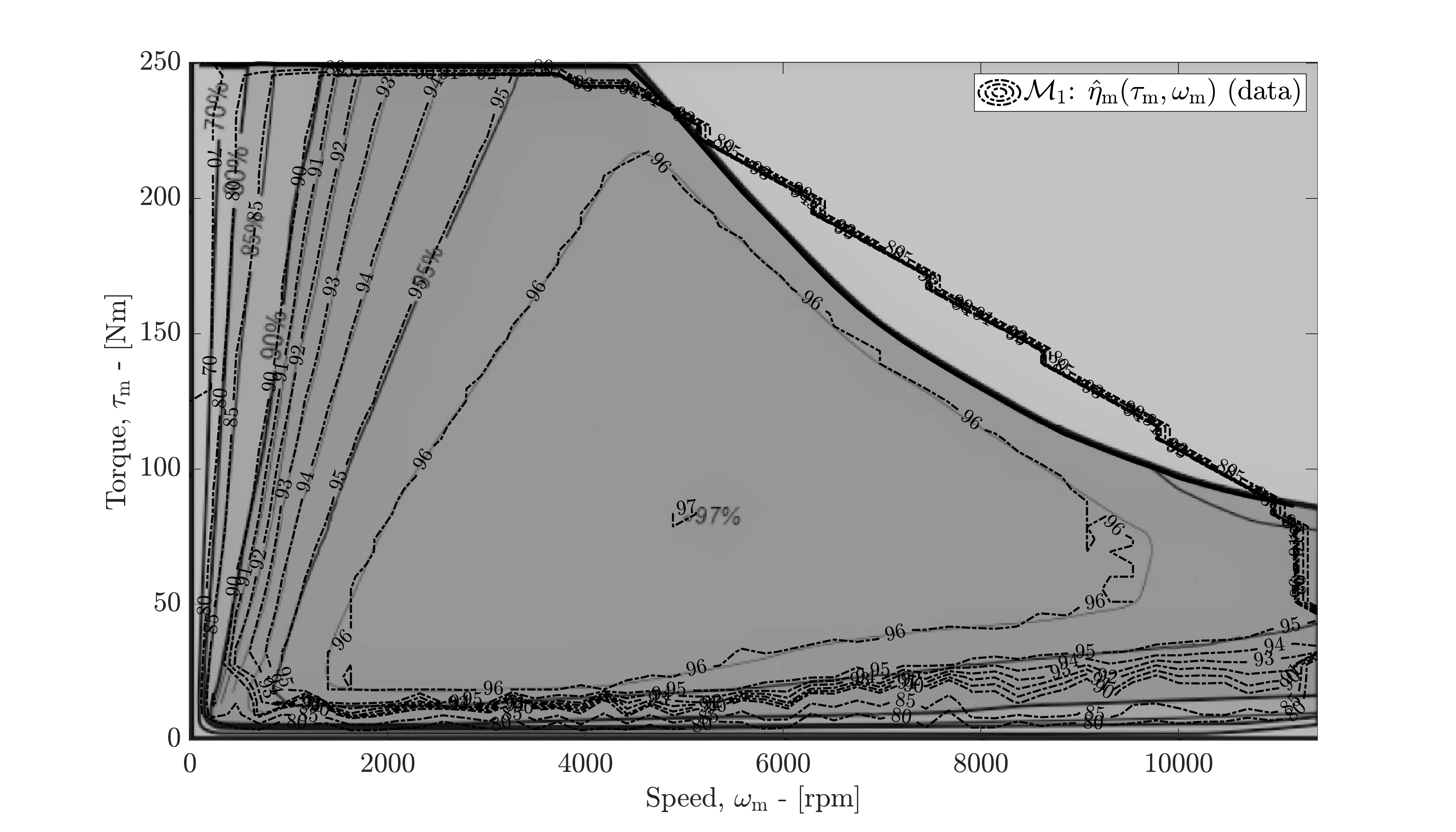}} \\
\subfloat[Data map $\mathcal{M}_1$ versus derived analytical map $\mathcal{M}_2$]{\includegraphics[clip = true, trim = 0cm 0cm 0cm 0cm,width= 1\columnwidth]{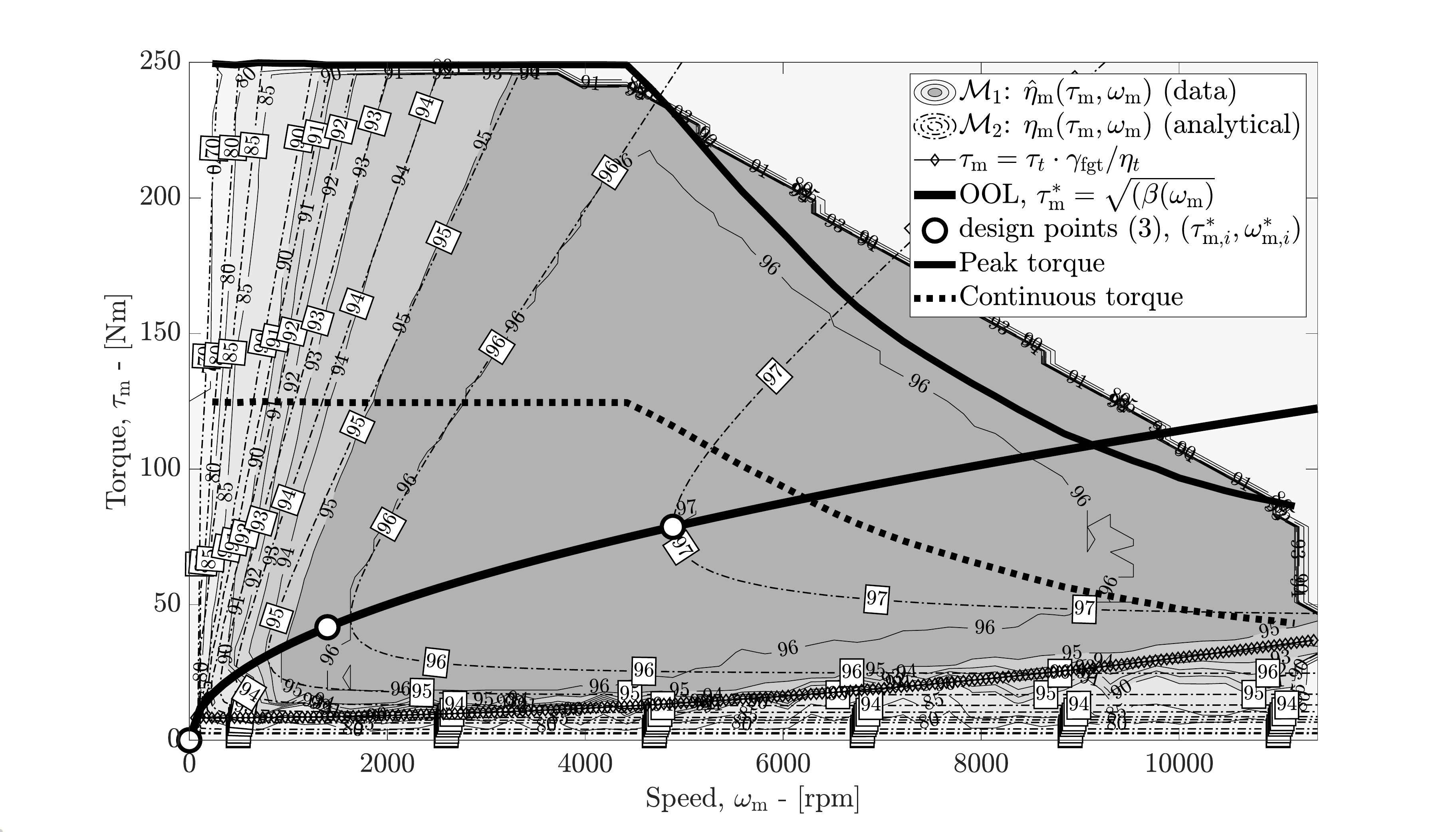}}
 \caption{Evaluation of the analytical model $\mathcal{M}_2$ using data of a high-efficient PMSM with a peak efficiency of 97\%.}\label{fig:fig2}
\end{figure}
The upper plot of Fig.~\ref{fig:fig2} shows the efficiency map of the original PMSM taken from~\cite{Mer14}, denoted as map model $\mathcal{M}_0$, together with the extracted data, denoted as $\mathcal{M}_1$. The overlay of the two maps show them to be in good agreement.
We use the data map $\mathcal{M}_1$ to fit the optimal operation line (OOL) with the three design points placed at the origin, at maximum efficiency and at a location chosen as
\begin{small}
\begin{equation}
(\underbrace{\tau_{\mathrm{m},2}^*(\omega_{\mathrm{m},2}^*)}_{\mbox{Eq.}\eqref{eq:opttrq}},\omega_{\mathrm{m},2}^*) = \arg \min_{\omega_{\mathrm{m},2} \in \omega_\mathrm{m}} \Delta \eta,
\end{equation}
\end{small}
\!\!in order to minimize the root mean squared error of the efficiency
\begin{small}
\begin{equation}
\Delta \eta = \left( \sum_{i=1}^{n}(\underbrace{\eta_\mathrm{m}^*(\omega_\mathrm{m}(i))}_{\mathcal{M}_2~:~\mbox{Eq.}\eqref{eq:effred}} - \underbrace{\hat{\eta}^*_\mathrm{m}(\overbrace{\tau_\mathrm{m}^*(\omega_\mathrm{m}(i))}^{\mbox{Eq.}\eqref{eq:opttrq}},\omega_\mathrm{m}(i))}_{\mathcal{M}_1~:~\mbox{data map}})^2 / n \right )^{\frac{1}{2}}.
\end{equation}
\end{small}
\!\!The location values for these design points are given by Table~\ref{tab:tab1}.
We use the resulting coefficients of the OOL listed in Table~\ref{tab:tab2} to reconstruct the analytical efficiency map $\mathcal{M}_2$, and compare it to $\mathcal{M}_1$ in the lower plot of Fig.~\ref{fig:fig2}, where also the peak and continuous torque values are shown for convenience.
Finally, we also map the stationary road load torque for the BMW~i3 equipped with a FGT, using the parameters listed in Table~\ref{tab:tab3} and assuming the vehicle mass to be sum of the weight of the empty car and a passenger as $m_\mathrm{v} = m_0 + m_\mathrm{p} = \unit[1195]{kg} + \unit[100]{kg} = \unit[1295]{kg}$, in line with the WLTC procedures.
\begin{table}[t]
\caption{Three design points on the optimal operation line}
\centering
\begin{tabular}{l|c|c|c|c} \toprule
point, $i$      & 1. & 2. & 3. & Units\\ \midrule
$\tau_{\mathrm{m},i}^*$   & 0 & 41.7 & 78.7 & (Nm) \\
$\omega_{\mathrm{m},i}^*$ & 0 & 1396 & 4886 & (rpm) \\
$\eta_{\mathrm{m},i}^*$   & 0 & 95.8 & 97 & (\%) \\ \bottomrule
\end{tabular}\label{tab:tab1}
\end{table}
\begin{table}[t]
\caption{Fit coefficients resulting in the RMSE $\Delta \eta = 0.44 \%$}
\centering
\begin{tabular}{l|l|l} \toprule
$d_{0 0} = 0$ & $d_{0 1} = 11.7843 $ & $d_{0 2} =  6.3048e-04 $ \\
$c_{0 0} = 0$ & $c_{0 1} = 0.5732$ & $c_{0 2} = 3.069e-05$ \\
$c_{1 1} = 0.0160$ & $c_{2 0} = 0.0487$ & \\ \bottomrule
\end{tabular}\label{tab:tab2}
\end{table}
\begin{table}[t]
\caption{Vehicle and powertrain parameters}
\centering
\small{
\begin{tabular}{lrl|lrl} \toprule
$m_0$ & 1195 & kg & $m_\mathrm{p}$ & 100 & kg\\
$A_\mathrm{f}$ & 2.38 & m$^2$ & $\bar{\tau}_\mathrm{m}$ / $\bar{\tau}_\mathrm{m,p}$ & 150 / 250 & Nm \\
$c_\mathrm{d}$ & 0.29 & - & $\omega_\mathrm{m,0}$ / $\bar{\omega}_\mathrm{m}$ & 4800 / 11400 & rpm \\
$c_\mathrm{r}$ & 1.74 & \% & $\bar{P}_\mathrm{m}$ / $\bar{P}_\mathrm{m,p}$ & 75 / 125 & kW\\
$r_\mathrm{w}$ & 0.350 & m &  $\lambda$ & 1.05 & -\\
$\eta_\mathrm{t}$ & 97 & \%  & $f_\mathrm{r}$ & 50/125 & - \\
$\kappa_\mathrm{R}$ & 0.55 & - & $\gamma$ & 9.665 \cite{Mer14} & - \\ \bottomrule
\end{tabular}}\label{tab:tab3}
\end{table}
\subsection{Energy Analysis: Two Transmission Technologies}
\begin{figure}[t]
  \centering
  \includegraphics[clip = true, trim = 3.5cm 8.5cm 3.5cm 8.5cm, width=1.0\columnwidth]{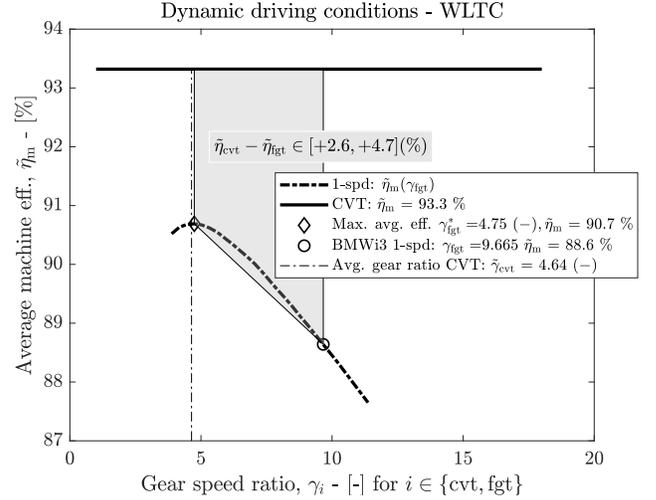}
  \caption{Average E-machine efficiency as a function of the gear speed ratio value for the FGT and the CVT.}\label{fig:fig3}
\end{figure}

We compute the average machine efficiency $\tilde{\eta}_\mathrm{m}$ over a reference drive cycle WLTC combining motoring and regenerative braking for a BMW~i3 equipped with a FGT or a CVT. The resulting values are 88.6\% and 93.3\%, respectively, with the CVT improving the operation efficiency of the machine by approximately +4.7\%.
This is further analysed by plotting the average efficiencies calculated over the WLTC as a function of the gear speed ratio values for the FGT and the CVT in Fig.~\ref{fig:fig3}.
In this diagram, the average machine efficiency is calculated for FGT as a function of its fixed gear speed ratio value, revealing potential for improvement: Specifically, if the FGT ratio was designed as $\gamma^*_{\mathrm{fgt}} = 4.75$, the efficiency would increase from 88.6\% to a maximum of 90.7\%.
In this case, the CVT would still outperform the FGT by at least 2.6\% without compromising the vehicle's top speed. The reader can verify that the average CVT speed ratio $\tilde{\gamma}_{\mathrm{cvt}} = 4.64$ comes close to this calculated constant speed ratio value $\gamma^*_{\mathrm{fgt}} = 4.75$ for the FGT.
In the next section, we will leverage our analytical approach to optimize the gear ratio design.
\subsection{Optimal Gear Ratio Design and Analysis}\label{sec:gear}
The optimal gear ratio as a function of the wheels' speed $\omega_\mathrm{t}$ and torque $\tau_\mathrm{t}$ can be derived substituting $\omega_\mathrm{m} = \omega_\mathrm{t} \, \gamma$ in~\eqref{eq:opttrq} as
\begin{small}
\begin{equation}
\tau_\mathrm{t} = \tau^*_\mathrm{m}(\omega_\mathrm{t} \, \gamma) \, \gamma \, \eta_\mathrm{t} = \sqrt{\beta(\omega_\mathrm{t} \ \gamma)} \, \gamma \, \eta_\mathrm{t},
\end{equation}
\end{small}
\!\!where $\eta_\mathrm{t}$ is the transmission efficiency.
From this, the optimal gear ratio $\gamma^*$ can be computed by solving
\begin{small}
\begin{align}
\left ( \frac{\tau_\mathrm{t}}{\eta_\mathrm{t}} \right )^2 - d_{0 0} \, \gamma^2 - d_{0 1} \, \omega_\mathrm{t} \, \gamma^3 - d_{0 2} \, \omega_\mathrm{t}^2 \, \gamma^4 = 0,
\end{align}
\end{small}
\!\!which results in the analytical solution for $\gamma^*(\tau_\mathrm{t}, \omega_\mathrm{t})$ given by \eqref{eq:optgear} with substitution parameters
\begin{small}
\begin{align}
\sigma_1 & = \begin{array}{l} 9 \, \sigma_2^{2/3} - 6 \, \sigma_5 \, \sigma_2^{1/3} + \sigma_5^2 - \frac{12 \, \kappa}{d_{0 2} \, \omega_\mathrm{t}} - \frac{9 \, d_{0 1}^4}{64 \, d_{0 2}^4 \, \omega_\mathrm{t}^4}  \\ + \frac{3 \, d_{0 0} \, d_{0 1}^2}{4 \, d_{0 2}^3 \omega_\mathrm{t}^4} \end{array} \\
\sigma_2 & = \frac{\sigma_6^2}{2} + \frac{4 \, \sigma_5 \, \sigma_4}{3} + \frac{\sqrt{3} \, \sigma_3}{18} + \frac{\sigma_5^3}{27} \\
\sigma_3 & = \sqrt{ \begin{array}{l} 256 \, \sigma_4^3 + 128 \, \sigma_5^2 \, \sigma_4^2 + 27 \, \sigma_6^4 + 4 \, \sigma_5^3 \, \sigma_6^2 + \\ 16 \, \sigma_5^4 \, \sigma_4 + 144 \, \sigma_5 \, \sigma_6^2 \, \sigma_4 \end{array} }  \\
\sigma_4 & = \frac{\kappa}{d_{0 2} \, \omega_\mathrm{t}} + \frac{3\, d_{0 1}^4}{256 \, d_{0 2}^4 \omega_\mathrm{t}^4} - \frac{d_{0 0} \, d_{0 1}^2}{16 \, d_{0 2}^3 \, \omega_\mathrm{t}^4} \\
\sigma_5 & = \frac{d_{0 0}}{d_{0 2} \, \omega_\mathrm{t}^2} - \frac{3 \, d_{0 1}^2}{8\, d_{0 2}^2 \, \omega_\mathrm{t}^2} \\
\sigma_6 & = \frac{d_{0 1}^3}{8 \, d_{0 2}^3 \, \omega_\mathrm{t}^3} - \frac{d_{0 0} \, d_{0 1}}{2\, d_{0 2}^2 \, \omega_\mathrm{t}^3} \\
\kappa & = \left ( \frac{\tau_\mathrm{t}}{\eta_\mathrm{t}} \right )^2.
\end{align}
\end{small}

\begin{figure}[t!]
  \centering
  \includegraphics[clip = true, trim = 2cm 6cm 2cm 5.5cm, width=1.0\columnwidth]{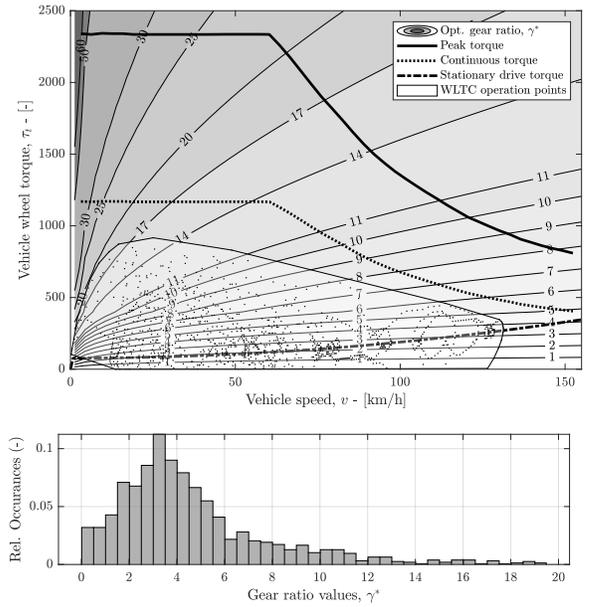}
  \caption{Optimal gear ratio maximizing the electric machine efficiency.}\label{fig:fig4}
\end{figure}

The proposed analytical models enable the construction of an efficiency map defining only three operation points on the optimal operation line.
What is more, the optimal speed ratio values for arbitrary wheel torque and speeds can be effectively analyzed as follows.
Fig.~\ref{fig:fig4} shows the optimal gear ratio resulting from~\eqref{eq:optgear}. Thereby, the minimum and maximum gear ratio values during stationary driving conditions from 0 to \unit[155]{km/h} are found to be:
\begin{small}
\begin{equation}
\gamma^*_{\mathrm{cvt}} \in \gamma_{\mathrm{cvt}} = [\underline{\gamma}_{\mathrm{cvt}},\overline{\gamma}_{\mathrm{cvt}}] = [2.92,4.40]
\end{equation}
\end{small}
\!\!at \unit[49]{km/h} and \unit[155]{km/h}, respectively.
In the same contour diagram, the WLTC operation points are plotted together with a histogram indicating the number of occurrences of each optimal gear ratio value. Thereby, it can be observed that more than \unit[80]{\%} of the gear ratio values are below 6, which could be covered by a single CVT, even tough the machine was designed by the OEM with an FGT.

\section{Conclusion}\label{sec:conc}
In this paper we leveraged analytical modeling tools to study the design of the transmission for electric vehicles.
In particular, we adopted an analytical approach to accurately describe an electric machine with just five parameters that can be found by identifying three operation points only, and validated it by fitting a BMW~i3 motor map.
Thereafter, we leveraged this model to compute the maximum-efficiency operational strategy for the transmission on any driving cycle, and applied it to compare two transmission technologies: a continuously variable transmission (CVT) and a fixed gear transmission (FGT).
We performed a numerical case study to show the effectiveness of our analytical approach, revealing that implementing a CVT instead of a FGT can significantly improve the energy efficiency of the electric machine, even if the OEM intentionally designed the machine for a FGT.
Therefore, we expect the benefits stemming from fully integrating the design concept of the electric machine and the transmission technology to be even higher.
In particular, our analytical model could enable to reverse the design process by optimizing the e-machine requirements as a function of the transmission design, and to ultimately integrate the design of both components in a joint optimization effort.
Ultimately, these preliminary results prompt us to carry out more detailed case-studies and pave the way to rethink the design of the motor and the transmission for electric automotive applications through an integrated-optimization rationale.

\section*{Acknowledgment}
The authors would like to thank Mr.\ W.\ van Harselaar for providing the stick diagrams of the shown transmission technologies, and Dr.\ I.\ New and Mr.\ F.\ Verbruggen for proofreading this paper.

\bibliographystyle{IEEEtran}
\bibliography{bibtex,../../../Bibliography/SML_papers,../../../Bibliography/main}

\newcommand{\noopsort}[1]{} \newcommand{\printfirst}[2]{#1}
  \newcommand{\singleletter}[1]{#1} \newcommand{\switchargs}[2]{#2#1}
\begin{thebibliography}{10}
\providecommand{\url}[1]{#1}
\csname url@samestyle\endcsname
\providecommand{\newblock}{\relax}
\providecommand{\bibinfo}[2]{#2}
\providecommand{\BIBentrySTDinterwordspacing}{\spaceskip=0pt\relax}
\providecommand{\BIBentryALTinterwordstretchfactor}{4}
\providecommand{\BIBentryALTinterwordspacing}{\spaceskip=\fontdimen2\font plus
\BIBentryALTinterwordstretchfactor\fontdimen3\font minus
  \fontdimen4\font\relax}
\providecommand{\BIBforeignlanguage}[2]{{%
\expandafter\ifx\csname l@#1\endcsname\relax
\typeout{** WARNING: IEEEtran.bst: No hyphenation pattern has been}%
\typeout{** loaded for the language `#1'. Using the pattern for}%
\typeout{** the default language instead.}%
\else
\language=\csname l@#1\endcsname
\fi
#2}}
\providecommand{\BIBdecl}{\relax}
\BIBdecl

\bibitem{Ver18}
F.~J.~R. {Verbruggen}, A.~E. {Hoekstra}, and T.~{Hofman}, ``Evaluation of the
  state-of-the-art of full-electric medium and heavy-duty trucks,'' in
  \emph{31th International Electric Vehicle Symposium \& Exhibition}, Kobe,
  Japan, 1-3 October 2018.

\bibitem{Ver19}
F.~J.~R. {Verbruggen}, V.~{Rangarajan}, and T.~{Hofman}, ``Powertrain design
  optimization for a battery electric heavy-duty truck,'' in \emph{2019
  American Control Conference (ACC)}, July 2019, pp. 1488--1493.

\bibitem{FIA}
FIA. History - the formula e story. Available online at
  \url{https://www.fiaformulae.com/en/discover/history}.

\bibitem{BorsboomFahdzyanaEtAl2020}
O.~Borsboom, C.~A. Fahdzyana, M.~Salazar, and H.~T., ``Time-optimal control
  strategies for electric race cars for different transmission technologies,''
  in \emph{{IEEE Vehicle Power and Propulsion Conference}}, 2020, submitted
  version available at \url{https://arxiv.org/abs/2005.07592}.

\bibitem{Verbruggen_2020}
\BIBentryALTinterwordspacing
F.~J.~R. Verbruggen, E.~Silvas, and T.~Hofman, ``Electric powertrain topology
  analysis and design for heavy-duty trucks,'' \emph{Energies}, vol.~13,
  no.~10, p. 2434, May 2020. [Online]. Available:
  \url{http://dx.doi.org/10.3390/en13102434}
\BIBentrySTDinterwordspacing

\bibitem{Zhu17}
Z.~Q. {Zhu}, W.~Q. {Chu}, and Y.~{Guan}, ``Quantitative comparison of
  electromagnetic performance of electrical machines for hevs/evs,'' \emph{CES
  Transactions on Electrical Machines and Systems}, vol.~1, no.~1, pp. 37--47,
  March 2017.

\bibitem{Choi13}
{Gilsu Choi} and T.~M. {Jahns}, ``Design of electric machines for electric
  vehicles based on driving schedules,'' in \emph{2013 International Electric
  Machines Drives Conference}, May 2013, pp. 54--61.

\bibitem{Lei17}
\BIBentryALTinterwordspacing
G.~Lei, J.~Zhu, Y.~Guo, C.~Liu, and B.~Ma, ``A review of design optimization
  methods for electrical machines,'' \emph{Energies}, vol.~10, no.~12, 2017.
  [Online]. Available: \url{https://www.mdpi.com/1996-1073/10/12/1962}
\BIBentrySTDinterwordspacing

\bibitem{Zuu2019}
\BIBentryALTinterwordspacing
M.~Zuurbier, C.~Fahdzyana, T.~Hofman, J.~Bao, and E.~Lomonova, ``Geometric
  optimization of variable flux reluctance machines for full electric
  vehicles,'' 2019, cited By 0. [Online]. Available:
  \url{https://www.scopus.com/inward/record.uri?eid=2-s2.0-85072305151&doi=10.1109%2fEVER.2019.8813611&partnerID=40&md5=db62a09ea176060f930bf944240b06f3}
\BIBentrySTDinterwordspacing

\bibitem{Sti15}
S.~{Stipetic}, D.~{Zarko}, and M.~{Popescu}, ``Scaling laws for synchronous
  permanent magnet machines,'' in \emph{2015 Tenth International Conference on
  Ecological Vehicles and Renewable Energies (EVER)}, March 2015, pp. 1--7.

\bibitem{Zhao16}
\BIBentryALTinterwordspacing
J.~Zhao and A.~Sciarretta, ``Design and control co-optimization for hybrid
  powertrains: Development of dedicated optimal energy management strategy,''
  \emph{IFAC-PapersOnLine}, vol.~49, no.~11, pp. 277 -- 284, 2016, 8th IFAC
  Symposium on Advances in Automotive Control AAC 2016. [Online]. Available:
  \url{http://www.sciencedirect.com/science/article/pii/S2405896316313635}
\BIBentrySTDinterwordspacing

\bibitem{Hup19}
I.~Hupkes, ``Variable drive ev: comfort solution for full electric vehicles,''
  in \emph{VDI 3rd Int. Conference CVT in automotive applications}, VDI, Ed.,
  Baden-Baden, Germany, 2019.

\bibitem{Gun13}
D.~{Gunji} and H.~{Fujimoto}, ``Efficiency analysis of powertrain with toroidal
  continuously variable transmission for electric vehicles,'' in \emph{IECON
  2013 - 39th Annual Conference of the IEEE Industrial Electronics Society},
  Nov 2013, pp. 6614--6619.

\bibitem{Rua16}
\BIBentryALTinterwordspacing
J.~Ruan, P.~Walker, and N.~Zhang, ``A comparative study energy consumption and
  costs of battery electric vehicle transmissions,'' \emph{Applied Energy},
  vol. 165, pp. 119 -- 134, 2016. [Online]. Available:
  \url{http://www.sciencedirect.com/science/article/pii/S0306261915016542}
\BIBentrySTDinterwordspacing

\bibitem{Gao15}
\BIBentryALTinterwordspacing
B.~Gao, Q.~Liang, Y.~Xiang, L.~Guo, and H.~Chen, ``Gear ratio optimization and
  shift control of 2-speed i-amt in electric vehicle,'' \emph{Mechanical
  Systems and Signal Processing}, vol. 50-51, pp. 615 -- 631, 2015. [Online].
  Available:
  \url{http://www.sciencedirect.com/science/article/pii/S0888327014002210}
\BIBentrySTDinterwordspacing

\bibitem{Guo17}
L.~{Guo}, B.~{Gao}, Q.~{Liu}, J.~{Tang}, and H.~{Chen}, ``On-line optimal
  control of the gearshift command for multispeed electric vehicles,''
  \emph{IEEE/ASME Transactions on Mechatronics}, vol.~22, no.~4, pp.
  1519--1530, Aug 2017.

\bibitem{Zhou19}
\BIBentryALTinterwordspacing
Z.~Zhou and M.~Huang, ``Regenerative braking algorithm for the electric vehicle
  with a seamless two-speed transmission,'' \emph{Proceedings of the
  Institution of Mechanical Engineers, Part D: Journal of Automobile
  Engineering}, vol. 233, no.~4, pp. 905--916, 2019. [Online]. Available:
  \url{https://doi.org/10.1177/0954407018755818}
\BIBentrySTDinterwordspacing

\bibitem{Rua18}
\BIBentryALTinterwordspacing
J.~Ruan, P.~D. Walker, J.~Wu, N.~Zhang, and B.~Zhang, ``Development of
  continuously variable transmission and multi-speed dual-clutch transmission
  for pure electric vehicle,'' \emph{Advances in Mechanical Engineering},
  vol.~10, no.~2, p. 1687814018758223, 2018. [Online]. Available:
  \url{https://doi.org/10.1177/1687814018758223}
\BIBentrySTDinterwordspacing

\bibitem{Tia18}
\BIBentryALTinterwordspacing
Y.~Tian, J.~Ruan, N.~Zhang, J.~Wu, and P.~Walker, ``Modelling and control of a
  novel two-speed transmission for electric vehicles,'' \emph{Mechanism and
  Machine Theory}, vol. 127, pp. 13 -- 32, 2018. [Online]. Available:
  \url{http://www.sciencedirect.com/science/article/pii/S0094114X18302234}
\BIBentrySTDinterwordspacing

\bibitem{Pin14}
S.~De~Pinto, G.~Mantriota, Bottiglione, Sorniotti, P.~Perlo, F.~Viotto, and
  P.~Camocardi, ``A four-wheel-drive fully electric vehicle layout with
  two-speed transmissions,'' 12 2014.

\bibitem{VerbruggenSalazarEtAl2019}
F.~J.~R. Verbruggen, M.~Salazar, M.~Pavone, and T.~Hofman, ``Joint design and
  control of electric vehicle propulsion systems,'' in \emph{{European Control
  Conference}}, 2020, in press.

\bibitem{Mer14}
\BIBentryALTinterwordspacing
J.~Merwerth, ``The hybrid-synchronous machine of the new {BMW} i3 \& i8,''
  online, 2014. [Online]. Available:
  \url{http://hybridfordonscentrum.se/wpcontent/uploads/2014/05/20140404_BMW.pdf}
\BIBentrySTDinterwordspacing

\bibitem{Mah17}
A.~{Mahmoudi}, W.~L. {Soong}, G.~{Pellegrino}, and E.~{Armando}, ``Loss
  function modeling of efficiency maps of electrical machines,'' \emph{IEEE
  Transactions on Industry Applications}, vol.~53, no.~5, pp. 4221--4231, Sep.
  2017.

\bibitem{Zha17}
J.~Zhao, ``Design and control co-optimization for advanced vehicle propulsion
  systems,'' PhD Thesis, Universit\'e Paris-Scalay IFP Energies nouvelles,
  October 2017.

\bibitem{HOF2017}
\BIBentryALTinterwordspacing
T.~Hofman and N.~Janssen, ``Integrated design optimization of the transmission
  system and vehicle control for electric vehicles,'' \emph{IFAC-PapersOnLine},
  vol.~50, no.~1, pp. 10\,072 -- 10\,077, 2017, 20th IFAC World Congress.
  [Online]. Available:
  \url{http://www.sciencedirect.com/science/article/pii/S2405896317323996}
\BIBentrySTDinterwordspacing

\bibitem{Pur13}
M.~{Pourabdollah}, N.~{Murgovski}, A.~{Grauers}, and B.~{Egardt}, ``Optimal
  sizing of a parallel phev powertrain,'' \emph{IEEE Transactions on Vehicular
  Technology}, vol.~62, no.~6, pp. 2469--2480, July 2013.

\end{thebibliography}

\vfill
\end{document}